\documentstyle[aps,titlepage]{revtex}
\title{Thermohydrodynamics for a van der Waals fluid}
\author{Pep Espa\~{n}ol }
\address{Dept. F\'{\i}sica Fundamental, UNED, 
Aptdo. 60141 E-28080, Madrid, Spain}
\date{23 February 2001}

\begin{document}
\maketitle

\begin{abstract}
Following a  cell-method due to van Kampen   for the calculation  of a
coarse-grained  free energy functional for  the  van der Waals gas, we
compute   a   corresponding  entropy   functional    from  microscopic
principles.  This entropy functional is  one of the building blocks of
the  recently developed GENERIC  framework.  This  framework allows to
obtain  in   a  thermodynamically    consistent    way  the  continuum
hydrodynamic equations    for a fluid  able  to   display liquid-vapor
coexistence.  Surface tension    appears naturally and the   resulting
model describes interfaces as diffuse regions, much in the same spirit
as   the gradient theory for  equilibrium  situations. We suggest that
using interfacial forces in the integral form obtained in the   microscopic   derivation  instead of    third  order
derivatives of the density field might represent an advantage from a
computational point of view.
\end{abstract}
% \bmulticol 
%\newpage
\section{Introduction}

Boiling of  water in a  pot, the formation of a  cloud, or the rise of
bubbles in a  pint of beer are  fascinating  everyday experiences that
involve       liquid-vapor    coexistence    in      non-trivial  flow
situations. Knowledge of the dynamics of this  phenomena is crucial in
less  pleasant situations like the   prediction and control of nuclear
accidents in refrigerated nuclear  reactors.   The complexity  of  the
problem requires the aid  of computer simulations  in order to extract
practical information.  Engineering conventional computational   fluid
mechanics approaches have  relied on effective hydrodynamic  equations
in  which the presence   of bubbles or   drops  is taken into  account
through  the introduction  of   void or vapor  fraction  which require
empirical    constitutive       equations   not      always  available
\cite{Hetsroni82}.   In   these much  coarse   grained approaches  the
detailed interface dynamics of bubbles and droplets is not available.

There  is a  great  recent  interest  in  the computer simulations  of
interfacial dynamics for  liquid-vapor coexistence.  Prove  of that is
the   large  variety  of  techniques  used    to  tackle the  problem:
Lattice-Boltzmann   method      \cite{Kato97,ThermohydroLB},    Direct
Montecarlo method  \cite{Alexander97}, Smoothed Particle Hydrodynamics
(SPH)\cite{Nugent00},    Finite  Difference  discretization         of
Navier-Stokes equations  \cite{Nadiga96},\cite{Gueyffier99},  and  the
Volume Of Fluid method \cite{Tsai96}.  Some of these techniques suffer
from ad-hoc assumptions, more prominently isothermal behavior.  Others
require quite specific model  technicalities in order to  get sensible
results or   the   specification of boundary  conditions    in complex
topologies.   It seems, therefore,    appropriate to review  here  the
theoretical foundation  of the hydrodynamic  equations for the van der
Waals fluid.

Although hydrodynamics equations for    the van der Waals  fluid  were
posed as early  as 1901 by Korteveg \cite{Korteveg01},\cite{Joseph94},
a clarification of the structure and  thermodynamic consistency of the
equations is convenient in order  to avoid potential problems.  In the
late  60's,  for example, a  variant  of the  theory was introduced by
Kawasaki  \cite{Kawasaki69},  following  the   ideas of   van   Kampen
\cite{vanKampen64}.     It can   be shown   that   Kawasaki  theory is
thermodynamically inconsistent.  In Kawasaki   theory, the long  range
attractive mean field potential  between molecules acts as an external
force in the momentum equation.  A gradient approximation of this mean
field potential leads to purely local  equations involving third order
spatial derivatives of the density field.   Such a theory was proposed
by  Felderhof who constructed  a  thermodynamically consistent set  of
hydrodynamic equations    for        a  van   der     Waals      fluid
\cite{Felderhof70}. However,  Felderhof theory deals with the inviscid
fluid and no dissipation nor fluctuations where included in his theory
so a generalization of his theory seems  appropriate. For an excellent
review of diffuse interface models described by third order derivative
terms  see Ref.   \cite{Anderson98}.   On    the  other  hand,     the
discretization of third order derivatives is subject to high numerical
errors.  From  a  computational point  of  view,  a treatment  of  the
diffuse interfaces though a   non-local integral term as in   Kawasaki
theory might represent an advantage worth exploring.

The   essential theoretical  tool  for   describing {\em  equilibrium}
properties of liquid-vapor interfaces is the Density Functional Theory
(DFT) \cite{Davis96}.  A well-known {\em local} realization of the DFT
known as the   Density  Gradient  Theory successfully describes    the
equilibrium properties of liquid-vapor interfaces like surface tension
and density profiles, and   the  correct relation  between   different
scaling exponents near the critical point \cite{Davis96}.  This is due
to the relatively smooth behavior of these interfaces as compared with
fluid-solid  interfaces for which  a non-local expression for the free
energy functional is needed \cite{Tarazona85}.   The work presented in
this paper can be understood  as a generalization of  the local DFT to
{\em  non-equilibrium}   situations  in  which   an entropy functional
instead of a free energy functional is used to  describe both flow and
thermal transport \cite{Tarazona99}.   

The strategy we follow is to derive the  {\em entropy} functional from
microscopic principles  following ideas of van  Kampen,  who derived a
coarse grained {\em free energy} functional for the  van der Waals gas
\cite{vanKampen64}.  This entropy   functional is one  of the building
blocks   of  a  recently    developed  framework  for  non-equilibrium
thermodynamics known as GENERIC \cite{Ottinger98}.  This two generator
formalism ensures thermodynamic  consistency of its dynamic equations.
Under a set of well-defined  assumptions, the GENERIC formalism can be
{\em derived} from first principles by  means of a projection operator
technique \cite{Ottinger98}.  It is therefore  not surprising that all
well-established dynamic equations   for  non-equilibrium systems  fit
into   the  GENERIC  formalism:   Relativistic   and  non-relativistic
hydrodynamics, kinetic theory of gases (the Boltzmann equation) and of
polymer systems in non-isothermal situations, chemical reactions, just
to  mention   a few,  have  the  GENERIC  structure \cite{Ottinger98}.
Significant new dynamical models for a number of complex systems, such
as       polymer/surface     interactions\cite{walls},          liquid
crystals\cite{lcp},  or polymer blends\cite{hco108}  have already been
obtained, and    even     new   cosmological    models  have      been
developed\cite{cosmo}.  In particular, the recognition  of the role of
the reversible dynamics  in the system evolution  equations has led to
interesting    new         conclusions       regarding         closure
approximations\cite{edw97}        and       polymer          reptation
models\cite{oet99,hco123}.  It seems, therefore, sensible to formulate
a thermohydrodynamic  theory for liquid-vapor coexistence according to
the GENERIC  framework.  The basic  bonus for  this  procedure is that
strict respect for  the  first and  second  law  of thermodynamics  is
guarded.

The paper is distributed as follows.  In Section \ref{sec1} we present
a brief summary of the GENERIC formalism and of the entropy functional
for   a van der Waals   fluid.   The entropy  functional is   computed
explicitly in the Appendix \ref{ap1}.  This appendix can be  regarded
as providing solid  microscopic  ground to the  plausible  assumptions
made by Felderhof when postulating the entropy functional.  In Section
\ref{hydro}  we propose the hydrodynamic equations  of a van der Waals
fluid according  to the  GENERIC  formalism and   compare it with  the
theory of Kawasaki.   In Section \ref{sec3}  the local gradient theory
is presented and  compared with  Felderhof's  theory. Some  concluding
remarks are   given in Section   \ref{discussion},  while some further
information is given in Appendix \ref{apmol}, and \ref{vdwhc}.

\section{The GENERIC framework and the entropy}
\label{sec1}
A brief summary of the GENERIC structure is presented  and we refer to
the original references for   further details  \cite{Ottinger98}.   The
first
essential step in the description of a system is  the selection of the
proper relevant variables $x$  used to describe the  system at a given
level of description.  The GENERIC dynamical equation for $x$ is

\begin{equation}
\dot{x}=L(x)\!\cdot\!\nabla E(x)
+M(x)\!\cdot\!\nabla S(x).
\label{gen1}
\end{equation}
The first term in the right hand  side produces the reversible part of
the dynamics  whereas    the  second term  is   responsible   for  the
irreversible dissipative  dynamics.  Here, $E(x),S(x)$  are the energy
and entropy  of the system expressed   in terms of  the variables $x$,
$\nabla$ is the gradient  operator  in $x$-space, and  $L(x),M(x)$ are
matrices that satisfy the following degeneracy requirements,

\begin{equation}
L\!\cdot\!\nabla S=0,\quad\quad\quad
M\!\cdot\!\nabla E=0.
\label{gen2}
\end{equation}
In  addition, $L$  is antisymmetric  (this guarantees  that energy  is
conserved, i.e.  the First  Law) and  satisfies the  stringent  Jacobi
property \cite{Ottinger98}.    $M$ is a  positive definite  symmetric
matrix (this guarantees that the  entropy is a nondecreasing  function
of  time,  i.e.  the Second Law).    If the system  presents dynamical
invariants  $I(x)$ different   from  the total  energy,  then  further
restrictions on the form of $L$ and $M$ are required
\begin{equation}
\nabla I\!\cdot\!
L\!\cdot\!\nabla E=0,\quad\quad\quad
\nabla I\!\cdot\!
M\!\cdot\!\nabla S=0.
\label{il}
\end{equation}
These conditions ensure that $d I/ dt =0$. The deterministic equations
(\ref{gen1})  are, actually, an     approximation in  which    thermal
fluctuations are neglected. If thermal fluctuations are not neglected,
the dynamics is   described  by a  Fokker-Planck equation  (FPE)  that
governs   the   probability   distribution   function $\rho=\rho(x,t)$
\cite{Ottinger98}. This FPE has as equilibrium solution the Einstein's
distribution function generalized to take into account the presence of
dynamical invariants $E_0,I_0$ \cite{Espanol92}, this is
\begin{equation}
\rho^{\rm eq}(x) = \delta(E(x)-E_0)\delta(I(x)-I_0)\exp \{ S(x)/k_B\},
\label{einst}
\end{equation}
where $k_B$ is Boltzmann's constant.

Our approach for constructing the entropy function for a liquid-vapor
system is to compute  microscopically $\rho^{\rm eq}(x)$  and, through
Eqn.  (\ref{einst}), identify the   entropy functional.  That is,   we
compute the  joint probability  that   an  extended simple  fluid   in
equilibrium has a  particular  realization of  the mass, momentum  and
internal energy density fields.  We follow a cell-method first used by
van  Kampen in  order to derive  the  equilibrium  probability  that a
simple  fluid has a particular  realization of the  mass density field
\cite{vanKampen64}.  The explicit details of the calculation are given
in  Appendix \ref{ap1}. Physical space  is divided in ${\cal M}$ cells
of volume ${\cal V}_\mu$ and the  mass $M_\mu$, momentum ${\bf P}_\mu$
and  internal energy   $\epsilon_\mu$ of   cell  $\mu$  in terms    of
microscopic coordinates of the  molecules is considered.  The internal
energy contains the  potential energy of interaction between particles
within  the  same  cell  {\em and}   the  energy  of  interaction with
particles of other  cells.  Provided that  the microscopic coordinates
are distributed according to the microcanonical ensemble we compute in
Appendix   \ref{ap1}  the  joint   probability $P[M,{\bf P},\epsilon]$
following similar steps as  in Ref.  \cite{vanKampen64}.  An essential
step in the derivation is the  assumption that the molecular potential
has  a repulsive   hard core  $\phi^{\rm   hc}(r)$ and   a long  range
attractive tail  $\phi^l(r)$.  The   interaction of particles  between
{\em  different} cells  is taken in  mean field  approximation and  it
involves only $\phi^l(r)$.  The final outcome is

\begin{eqnarray}
P[M,{\bf P},{\epsilon}]
&=&
\frac{1}{\Omega_0}\delta\left(\sum^{\cal M}_\mu M_\mu - M_0\right)
\delta\left(\sum_\mu^{\cal M}{\bf P}_\mu\right)
\label{PNPE}\\
&\times&
\delta\left(\sum_\mu^{\cal M}
\left(\frac{{\bf P}^2_\mu}{2M_\mu}+\epsilon_\mu\right)-E_0\right)
\exp \left\{\frac{S[M,{\epsilon}]}{k_B}\right\}.
\nonumber
\end{eqnarray}
We   recognize the  dynamical invariants  (total  mass $M_0$, momentum
${\bf  P}_0={\bf  0}$ and   energy   $E_0$) in  the conserving   delta
functions.  The entropy functional is given by
\begin{equation}
S[M,{\epsilon}]
=\sum_\mu S^{\rm hc}(\epsilon_\mu-\overline{\phi}_\mu,M_\mu,{\cal V}_\mu).
\label{mesent}
\end{equation}
Here,  $S^{\rm hc}(\epsilon,M,V)$   is  the  entropy  function  of  an
isolated system  of $N=M/m$ molecules ($m$  is the mass of a molecule)
interacting with {\em only} the hard core potential $\phi^{\rm hc}(r)$
in a volume $V$. Note  the non-trivial appearance of the  mean
field interaction energy $\overline{\phi}_\mu$, which is defined by
\begin{equation}
\overline{\phi}_\mu =\frac{1}{2m^2}\sum_\nu'\phi^l(R_{\mu\nu})
M_{\mu}M_{\nu},
\label{C10}
\end{equation}
where $R_{\mu\nu}$ is the distance between cell  centers and the prime
denotes $\mu\neq\nu$. This mean field  potential involves only the long
ranged part  of the molecular  potential.  Integration of (\ref{PNPE})
with  respect  to    $N,\epsilon$  and  using   a  steepest    descent
approximation (see    Appendix  \ref{apmol}) leads  to   van  Kampen's
expression,

\begin{equation}
P[M]=\frac{1}{Z_0}
\exp \left\{-\beta F[M]\right\}\delta\left(\sum^{\cal M}_\mu M_\mu - M_0\right),
\label{pn}
\end{equation}
where the coarse-grained free energy functional is given by
\begin{equation}
F[M]=\sum_\mu^{\cal M}
F^{\rm hc}(\beta,M_\mu,{\cal V}_\mu)+\beta\overline{\phi}_\mu.
\label{FN}
\end{equation}
Here, $F^{\rm hc}(\beta,M_\mu,{\cal  V}_\mu)$ is the  free energy of a
system of  hard core particles  at temperature $\beta^{-1}/k_B= 2 K^*/
DM$ ($K^*$ is the most  probable value of  the  kinetic energy in  the
system).  Both Eqns.    (\ref{PNPE}) and (\ref{pn})  admit a continuum
notation provided  that the cells are  large enough  in order that the
entropy and free energy are first order functions of their arguments
\begin{eqnarray}
S[\rho_{\bf r},{\epsilon}_{\bf r}]
&=&\int d{\bf r} s^{\rm hc}
(\epsilon_{\bf r}-\overline{\phi}_{\bf r},\rho_{\bf r}),
\label{sf}
\\
F[\rho_{\bf r}]&=&\int d{\bf r} 
\left(f^{\rm hc}(\rho_{\bf r},\beta)+\overline{\phi}_{\bf r}\right),
\label{ff}
\end{eqnarray}
where      $\epsilon_{\bf    r}=\epsilon_{\mu}/{\cal          V}_\mu$,
$\overline{\phi}_{\bf r}=\overline{\phi}_\mu/{\cal V}_\mu$, $\rho_{\bf
r}=M_{\mu}/{\cal       V}_\mu$,        and  $s^{\rm  hc}(\epsilon_{\bf
r}-\overline{\phi}_{\bf r},\rho_{\bf r})$ is the entropy {\em density}
of a hard core system,  while $f^{\rm hc}(\rho_{\bf r},\beta)$ is  the
corresponding free energy  density.  The continuum form for
the mean field potential energy (\ref{C10}) is

\begin{equation}
\overline{\phi}_{\bf r}= 
\rho_{\bf r}\int d{\bf r}' \frac{\phi^l(|{\bf r}-{\bf r}'|)}{2 m^2}
\rho_{{\bf r}'}
\label{appot0}
\end{equation}

The main benefit of the  microscopic derivation presented in  Appendix
\ref{apmol}  is the clear separation at  the macroscopic  level of the
effects of the short and long ranged part of the microscopic potential,
in particular the realization  that the entropy  of the full system is
given   directly in terms   of the entropy  of  the  hard core system,
suitably evaluated     at    the     corrected    internal     energy,
Eqns. (\ref{mesent}) or (\ref{sf}).

We now consider  the particular situation  in which the  density field
 varies slowly in the scale of the range of $\phi^l(r)$. In this case,
 we can Taylor expand $\rho_{{\bf r}'}$ around ${\bf r}$ and retain up
 to second order in gradients in order to get

\begin{equation}
\overline{\phi}_{\bf r} \approx -a\rho^2_{\bf r}-\frac{c}{2}\rho_{\bf r}
\nabla^2\rho_{\bf r},
\label{appot}
\end{equation}
where we introduced $a,c>0$ through
\begin{equation}
a =  -\int d{\bf r}\frac{\phi^l(r) }{2m^2},
\quad\quad
c = -\int d{\bf r}r^2\frac{\phi^l(r)}{3m^2}.
\end{equation}
Substitution of (\ref{appot}) into (\ref{ff}) leads to
\begin{equation}
F[\rho_{\bf r}]=\int d{\bf r} 
\left(f^{\rm hc}(\rho_{\bf r},\beta)
-a\rho^2+\frac{c}{2}(\nabla \rho_{\bf r})^2\right),
\label{fok}
\end{equation}
which is the usual local   gradient  density expression for the   free
energy functional, also known  as the Cahn-Hilliard or Ginzburg-Landau
free  energy   functional for  liquid   surfaces  \cite{Davis96}.  The
corresponding entropy functional (\ref{sf}),  thus, has been  computed
at the same level of approximation.   The coefficient $c$ provides the
overall magnitude of the surface tension coefficient \cite{Davis96}.

The energy function given by
\begin{equation}
E(M,{\bf P},\epsilon)=\sum^{\cal M}_\mu
\left(\frac{{\bf P}_\mu^2}{2M_\mu}+\epsilon_\mu\right),
\label{Edisc}
\end{equation}
can also be written in continuum form as an energy functional

\begin{equation}
E[\rho_{\bf r},{\bf g}_{\bf r},\epsilon_{\bf r}]
= 
\int \left[\frac{1}{2}\frac{ {\bf g}_{\bf r}^2}{\rho _{\bf r}}
+ \epsilon_{\bf r}\right] d{\bf r}.
\label{Econt}
\end{equation}
where ${\bf g}_{\bf  r}={\bf P}_{\mu}/{\cal V}_{\mu}$ is  the momentum
density field.

\section{Hydrodynamic equations for a van der Waals fluid}
\label{hydro}

In this section, we construct the hydrodynamic equations for a van der
Waals fluid following the GENERIC formalism.

Due to the form  in which the  internal energy appears in the entropy,
it proves  convenient to use  as the proper  hydrodynamic variable the
``intrinsic''    internal  energy $\epsilon_{\bf    r}^i=\epsilon_{\bf
r}-\overline{\phi}_{\bf   r}$  which   involves  the   interaction  of
molecules through  the hard core  part only.  The energy (\ref{Econt})
and the  entropy (\ref{sf}) expressed  in terms of  these hydrodynamic
variables are then

\begin{eqnarray}
E[\rho,{\bf g},\epsilon^i]&=& 
\int \left[\frac{1}{2}\frac{ {\bf g}_{\bf r}^2}{\rho _{\bf r}}
+ \epsilon^i _{\bf r}+\overline{\phi}_{\bf r}\right] d{\bf r},
\label{energy}
\nonumber\\
S[\rho,{\bf g},\epsilon^i]&=& 
\int s^{hc}(\rho _{\bf r}, \epsilon^i _{\bf r}) d{\bf r}.
\label{entropy}
\end{eqnarray}
  We   have now the two  basic  building blocks  for the
GENERIC formulation    of the dynamic    equations  (\ref{gen1}).  The
derivatives of  the energy and  entropy  with respect  to the relevant
variables  become  functional  derivatives    with respect  to     the
hydrodynamic variables, this is,

\begin{equation}
\frac{\partial E}{\partial x}= 
\left( \begin{array}{c}
\frac{\delta  E}{\delta \rho_{\bf r}} \\
\\
\frac{\delta E}{\delta {\bf g}_{\bf r}}  \\
\\
\frac{\delta E}{\delta {\epsilon}^i_{\bf r}}  \\
\end{array} \right) 
=
\left( \begin{array}{c}
-\frac{1}{2}  {\bf v} _{\bf r}^2
-\frac{2\overline{\phi} _{\bf r}}{\rho_{\bf r}}\\
\\
{\bf v} _{\bf r}\\
\\
1
\end{array} \right),
\label{nablaE}
\end{equation}
and
\begin{equation}
\frac{\partial S}{\partial x}=
\left( \begin{array}{c}
\frac{\delta S}{\delta \rho_{\bf r}}  \\
\\
\frac{\delta S}{\delta {\bf g}_{\bf r}}\\
\\
\frac{\delta S}{\delta {\epsilon}^i_{\bf r}} \\
\end{array} \right) 
=  \left( \begin{array}{c}
-\frac{\mu^{\rm hc}(\rho_{\bf r},\epsilon _{\bf r}^i)}
{T^{\rm hc}(\rho_{\bf r},\epsilon _{\bf r}^i)}
\\
\\
0\\
\\
\frac{1}{T^{\rm hc}(\rho_{\bf r},\epsilon _{\bf r}^i)}
\end{array} \right). 
\label{nablaS}
\end{equation}
The velocity  field is ${\bf  v  }_{\bf r}={\bf  g }_{\bf r}/\rho_{\bf
r}$, and  the  local temperature $T^{\rm  hc}$ and  chemical potential
$\mu^{\rm hc}$ per unit mass are 
\begin{equation}
T^{\rm hc}_{\bf r}=\left (\frac{\partial s^{\rm hc}}
{\partial \epsilon^i_{\bf r}}\right)^{-1},
\quad\quad\quad
\frac{\mu^{\rm hc}_{\bf r}}{T^{\rm hc}_{\bf r}}=
-\frac{\partial s^{\rm hc}}
{\partial \rho_{\bf r}}.
\label{intensive}
\end{equation}
Note that these equations of state are those of the hard core system.
By focusing on the reversible part  of the dynamics, we must construct
now  the matrix    $L$. A reasonable  proposal  is   to  use the  same
expression for  $L$ as   in the case  of   hydrodynamics of  a  simple
fluid.  Note  that  if  we  neglect the  interfacial  potential energy
$\overline{\phi}_{\bf  r}$ we should  recover  the hydrodynamics of  a
simple hard  core    fluid.  We  propose, therefore,  for   the matrix
$L\rightarrow    L_{{\bf     r}{\bf  r}'}    $    the   following  one
\cite{Ottinger98}

\begin{equation}
\left( \begin{array}{ccc}
0 & \rho_{{\bf r}'}\nabla'\delta_{{\bf r}{\bf r}'}& 0\\
&&\\
\rho _{\bf r}\nabla'\delta_{{\bf r}{\bf r}'}
& {\bf g}_{{\bf r}'}\nabla'\delta_{{\bf r}{\bf r}'}
+\nabla'\delta_{{\bf r}{\bf r}'} {\bf g}_{\bf r} 
&  \epsilon_{\bf r}^i\nabla'\delta_{{\bf r}{\bf r}'}
+ P^{\rm hc}_{{\bf r}'}\nabla'\delta_{{\bf r}{\bf r}'}
\\
&&\\
0 & \epsilon_{{\bf r}'}^i\nabla'\delta_{{\bf r}{\bf r}'}
+P^{\rm hc}_{\bf r}\nabla'\delta_{{\bf r}{\bf r}'}
& 0
\end{array} \right).
\label{Lrr}
\end{equation}
Here,   $\nabla'=\partial/\partial{\bf  r}'$ and  $\delta_{{\bf r}{\bf
r}'}=\delta({\bf r}-{\bf r}')$  is  Dirac's  delta  function,  $P^{\rm
hc}_{\bf  r}=P^{\rm   hc}(\rho_{\bf r},\epsilon_{\bf   r}^i)$  is  the
pressure field as a function of the mass and intrinsic internal energy
densities.

As has been shown in Ref.  \cite{Ottinger98}, the matrix $L$ satisfies
the condition  $L\nabla S=0$ because  of  the Gibbs-Duhem relationship
(for the hard core system).  The fulfillment of $L\nabla  S =0$ is the
basic  reason  for using $P^{\rm  hc}_{\bf  r}$  in  Eqn.  (\ref{Lrr})
instead of any other possible  expression for the pressure. The matrix
$L$  also satisfies the  stringent Jacobi  identity \cite{Ottinger98}.
With the explicit form (\ref{Lrr}) for the $L$ matrix and the gradient
of the energy in Eqn.  (\ref{nablaE}) the  reversible part $L\nabla E$
of the equations of motion for  the hydrodynamic variables are readily
constructed

\begin{eqnarray}
\partial_t \rho_{\bf r} 
&=& -{\bf \nabla}\rho_{\bf r} {\bf v}_{\bf r},
\nonumber\\
\partial_t {\bf g}_{\bf r} 
&=& -{\bf \nabla} {\bf v}_{\bf r} {\bf g}_{\bf r} 
-\nabla P^{\rm hc}(\epsilon_{\bf r}^i,\rho_{\bf r})
-\rho_{\bf r}\nabla\frac{2\overline{\phi} _{\bf r}}{\rho_{\bf r}},
\nonumber\\
\partial_t \epsilon^i_{\bf r} 
&=& -{\bf \nabla}\epsilon_{\bf r}^i{\bf v}_{\bf r}
-P^{\rm hc}(\epsilon_{\bf r}^i,\rho_{\bf r}){\bf \nabla}{\bf v}_{\bf r}.
\label{revhyd}
\end{eqnarray}
Eqns.  (\ref{revhyd})  are  identical to  the  Euler  equations  of an
inviscid  (hard  cored) fluid except   for the additional  term in the
momentum equation that  involves the long  range part of the molecular
potential.

%\emulticol 

Concerning the irreversible  part of the dynamics  $M\nabla
S$, we observe that the same matrix $M\rightarrow M_{{\bf r}{\bf r}'}$
that corresponds to  the  usual  hydrodynamic equations is   perfectly
adequate. For simplicity,  we assume  that there  is no extra
dissipation due to the presence of interfaces. The matrix $M$ is
thus given by \cite{Ottinger98}
\begin{equation}
\left( \begin{array}{ccc}
0 & 0 & 0\\
&&\\
0
&(\nabla \nabla'+{\bf 1}\nabla\!\cdot\!\nabla')\eta_{\bf r} T^{\rm hc}_{\bf r}
\delta_{{\bf r}{\bf r}'}
+2\nabla\nabla'\hat{\kappa}_{\bf r} T^{\rm hc}_{\bf r}\delta_{{\bf r}{\bf r}'}
&\nabla\eta_{\bf r} T^{\rm hc}_{\bf r}\dot{\gamma}\delta_{{\bf r}{\bf r}'}
+\nabla \hat{\kappa}_{\bf r}T^{\rm hc}_{\bf r}{\rm tr}
\dot{\gamma}\delta_{{\bf r}{\bf r}'}
\\
&&\\
0 & \nabla'\eta_{\bf r} T^{\rm hc}_{\bf r}\dot{\gamma}\delta_{{\bf r}{\bf r}'}
+\nabla' \hat{\kappa}_{\bf r}T^{\rm hc}_{\bf r}{\rm tr}
\dot{\gamma}\delta_{{\bf r}{\bf r}'}
& 
\frac{1}{2}\eta_{\bf r} T^{\rm hc}_{\bf r}\dot{\gamma}
:\dot{\gamma}\delta_{{\bf r}{\bf r}'}
+\nabla\nabla'\lambda_{\bf r} T^{\rm hc 2}_{\bf r}\delta_{{\bf r}{\bf r}'}
+\frac{1}{2}\hat{\kappa}_{\bf r} T^{\rm hc}_{\bf r}({\rm tr}
\dot{\gamma})^2\delta_{{\bf r}{\bf r}'}
\end{array} \right),
\label{Mrr}
\end{equation}
%\bmulticol 

where  $\dot{\gamma}=\nabla  {\bf  v}_{\bf  r}+(\nabla {\bf
v}_{\bf   r})^T$ is the symmetrized   velocity  gradient tensor.  This
matrix   $M$ satisfies the degeneracy $M\nabla    E=0$ as in the usual
hydrodynamic   case \cite{Ottinger98},  as   the  only modification of
$\nabla  E$ is in  the  term in the   density  component.  We  remark,
however, that the  transport coefficients (shear viscosity  $\eta_{\bf
r}$,    bulk viscosity $\kappa_{\bf r}=\hat{\kappa}_{\bf r}+2\eta_{\bf
r}/3$ and  thermal conductivity $\lambda_{\bf  r}$) that appear in $M$
should be  taken as state dependent  and  are expected to  be strongly
varying functions of the density field, in order to encompass the fact
that vapor and liquid have very different  transport properties.  This
is the reason  for its space dependence displayed  as a subindex ${\bf
r}$. The presence of $T^{\rm hc}_{\bf r}$ in Eqn.  (\ref{Mrr}) instead
of other possible choices for the temperature, is dictated by the fact
that $M\nabla S$ should  produce  the usual  dissipative terms  in the
Navier-Stokes  equation (note that $\nabla  S$  in Eqn. (\ref{nablaS})
involves $T^{\rm hc}_{\bf r}$).

The final set of hydrodynamic equations for the thermohydrodynamics of
the van der Waals fluid are

\begin{eqnarray}
\partial_t \rho_{\bf r} 
&=& -{\bf \nabla}\rho_{\bf r} {\bf v}_{\bf r},
\nonumber\\
\partial_t {\bf g}_{\bf r} 
&=& -{\bf \nabla} \!\cdot\! {\bf v}_{\bf r} {\bf g}_{\bf r} 
-\nabla P^{\rm hc}(\epsilon_{\bf r}^i,\rho_{\bf r})
-\nabla \!\cdot\! {\bf \tau}+\rho_{\bf r}\overline{\bf F}_{\bf r},
\nonumber\\
\partial_t \epsilon^i_{\bf r} 
&=& -{\bf \nabla}\epsilon_{\bf r}^i{\bf v}_{\bf r}
-P^{\rm hc}(\epsilon_{\bf r}^i,\rho_{\bf r}){\bf \nabla}\!\cdot\! 
{\bf v}_{\bf r}
-\nabla\!\cdot\!{\bf q} -{\bf \tau}:\nabla{\bf v}_{\bf r},
\label{fin}
\end{eqnarray}
where the constitutive equations for the viscous stress ${\bf \tau}$
and heat flux are given by

\begin{eqnarray}
{\bf \tau}_{\bf r} &= &-\eta_{\bf r} \left[\nabla {\bf v}_{\bf r}
+ \nabla {\bf v}_{\bf  r}^T\right]-(\kappa_{\bf r}- 2\eta_{\bf r}/3)
(\nabla\!\cdot\!{\bf v}_{\bf r}){\bf 1}, \label{ta}
\nonumber\\
{\bf q}_{\bf r}&=& -\lambda_{\bf r}\nabla T^{\rm hc}_{\bf r}.
\label{qa}
\end{eqnarray}
and the mean field force is given by
\begin{equation}
\overline{\bf F}_{\bf r}
=-\nabla_{\bf r}
\frac{1}{m^2}\int d{\bf r}'\overline{\phi}^l(|{\bf r}-{\bf r}'|)
\rho_{{\bf r}'}
\end{equation}

Equations   (\ref{fin})  are    deterministic   equations  in    which
thermodynamic fluctuations   are neglected. Thermodynamic fluctuations
are very easily included in the GENERIC formalism. One simply needs to
take the square  root  in matrix sense of   the $M$ matrix,   and this
provides the amplitude of the  noises \cite{Ottinger98}, in accordance
with  the Fluctuation-Dissipation theorem.  Because  the matrix $M$ in
Eqn. (\ref{Mrr}) has the same structure of that of  a one phase fluid,
we can  advance that the final  fluctuating  equations are obtained by
simply adding to the stress tensor $\tau_{\bf  r}$ and heat flux ${\bf
q}_{\bf r}$ a random counterparts given by \cite{Espanol98}

\begin{eqnarray}
\tilde{\bf \tau}_{\bf r} &=& \sqrt{2T^{\rm hc}_{\bf r}\eta_{\bf r}}
(\tilde{\bf \sigma}_{\bf r}(t)- 
{\rm tr}[\tilde{\bf \sigma}_{\bf r}(t)]{\bf 1})
+ \sqrt{3T^{\rm hc}_{\bf r}\kappa_{\bf r}}
{\rm tr}[\tilde{\bf \sigma}_{\bf r}(t)]{\bf 1}
\nonumber\\
\tilde{\bf q}_{\bf r}&=&T^{\rm hc}_{\bf r}
\sqrt{2\lambda_{\bf r}}\tilde{\bf \zeta}_{\bf r}(t),
\end{eqnarray}
where $\tilde{\bf \sigma}_{\bf r}(t)$  is a matrix of delta correlated
white noises and $\tilde{\bf \zeta}_{\bf  r}(t)$ is a vector of  delta
correlated white noises  (in space and time).  Note  that  the form in
which thermal fluctuations appear  is similar to the approach proposed
by Landau and Lifshitz \cite{Landau59}.  However, one should note that
in counterdistintion to their approach,  here the random stress tensor
and heat flux  are {\em  multiplicative} noises  which {\em are   not}
Gaussian.   What    is  Gaussian are   the  white   noises $\tilde{\bf
\sigma}_{\bf    r}(t),\tilde{\bf  \zeta}_{\bf     r}(t)$.  The  strong
dependence of  the transport coefficients  on  the density due to  the
phase change  has a direct  effect   on the amplitude of  the  thermal
fluctuations. 

Associated  to    the stochastic  differential   equations  describing
hydrodynamic fluctuating variables  there is a mathematical equivalent
functional  Fokker-Planck equation \cite{Espanol98}.     The
GENERIC   structure  ensures  that  the equilibrium   solution of this
Fokker-Planck equation   is    given by  the    continuum   version of
Eqn. (\ref{PNPE}), as it should. 

Kawasaki proposed a set of hydrodynamic equations for  a van der Waals
fluid on  intuitive  grounds \cite{Kawasaki69}.  Both   the continuity
equation and the momentum  balance equation in (\ref{revhyd}) coincide
with Kawasaki's postulated equations.  On the  other hand, he uses the
entropy per  unit  mass $\hat{s}^{\rm   hc}_{\bf  r}= s^{\rm  hc}_{\bf
r}/\rho_{\bf  r}$    instead  of   the   intrinsic   internal   energy
$\epsilon^i_{\bf  r}$.   A  standard    calculation which   uses   the
definitions (\ref{intensive}) and the Euler equation $T^{\rm hc}s^{\rm
hc}+\mu^{\rm  hc}\rho    -P^{\rm  hc}-\epsilon^i=0$  shows    that our
equations  (\ref{revhyd})  imply  $\partial_t   \hat{s}^{\rm  hc}+{\bf
v}_{\bf r}  \!\cdot\!\nabla \hat{s}^{\rm  hc}=0$.  This  is consistent
with the  fact that  the reversible  part of  the dynamics should  not
produce  an  increase of the entropy  in  the system.  In contrast, in
Kawasaki treatment the  reversible part of  the substantial derivative
of the entropy  per unit mass  is equated to $\rho_{\bf r}{\bf v}_{\bf
r}\!\cdot\!\nabla   (\overline{\phi}_{\bf  r}/\rho_{\bf r})$ (in   our
notation).  It is apparent that the addition of this term violates the
conservation of  the energy (\ref{energy}) and   should not be present
for thermodynamic consistency.
%\newpage
\section{Gradient approximation}
\label{sec3}
In this section  we consider the gradient  approximation (\ref{appot})
as applied to the obtained  equations (\ref{fin}).  The last two terms
of the momentum equation can be re-arranged in the form
\begin{equation}
-\nabla P^{\rm hc}_{\bf r} -\rho_{\bf r} 
\nabla\frac{\overline{\phi}_{\bf r}}{\rho_{\bf r}}
=-\nabla \overline{P}_{\bf r}
+c\rho_{\bf r}\nabla \nabla^2\rho_{\bf r},
\label{pressure}
\end{equation}
where we have introduced

\begin{equation}
\overline{P}_{\bf r}=P^{\rm hc}(\rho_{\bf r},\epsilon_{\bf r}^i)
-a\rho_{\bf r}^2.
\label{pvdw}
\end{equation}
The momentum equation thus becomes

\begin{equation}
\partial_t {\bf g}_{\bf r} 
= -{\bf \nabla} {\bf v}_{\bf r} {\bf g}_{\bf r} 
-\nabla \overline{P}(\epsilon_{\bf r}^i,\rho_{\bf r})
+c\rho_{\bf r}\nabla\nabla^2\rho_{\bf r}.
\label{mom1}
\end{equation}

It is illuminating to use as the set of 
hydrodynamic variables  $\rho_{\bf  r},{\bf  g}_{\bf   r},  u_{\bf r}$
instead of $\rho_{\bf r},{\bf  g}_{\bf r}, \epsilon^i_{\bf  r}$. Here,
the local internal energy is defined by
\begin{equation}
u_{\bf r} = \epsilon^i_{\bf r}-a \rho_{\bf r}^2,
\label{u}
\end{equation}
which,   from a microscopic point   of  view, represents the potential
energy due to local interactions, that is, hard core interactions plus
the mean field attraction ``within'' the small volume at ${\bf r}$.  By
taking the time derivative  of  (\ref{u}) and using  the  hydrodynamic
equations (\ref{fin}), a standard calculation leads to

\begin{eqnarray}
\partial_t \rho_{\bf r} 
&=& -{\bf \nabla}\rho_{\bf r} {\bf v}_{\bf r},
\nonumber\\
\partial_t {\bf g}_{\bf r} 
&=& -{\bf \nabla} {\bf v}_{\bf r} {\bf g}_{\bf r} 
-\nabla P^{\rm vdW}(u_{\bf r},\rho_{\bf r})
+c\rho_{\bf r}\nabla\nabla^2\rho_{\bf r}
-\nabla \!\cdot\! {\bf \tau},
\nonumber\\
\partial_t u_{\bf r} 
&=& -{\bf \nabla}u_{\bf r}{\bf v}_{\bf r}
-P^{\rm vdW}(u_{\bf r},\rho_{\bf r}){\bf \nabla}\!\cdot\! {\bf v}_{\bf r}
-\nabla\!\cdot\!{\bf q} -{\bf \tau}:\nabla{\bf v}_{\bf r},
\label{fin2}
\end{eqnarray}
where the heat flux is given by
\begin{equation}
{\bf q}_{\bf r} = -\lambda_{\bf r}\nabla T^{\rm vdW}(u_{\bf r},\rho_{\bf r}).
\label{qv}
\end{equation}
Here, $P^{\rm vdW},T^{\rm vdW}$ are the two equations of state for the
van   der    Waals fluid  (see   Appendix  \ref{vdwhc}).  In obtaining
(\ref{fin2}) and  (\ref{qv}), use  has   been made  of the  identities
(\ref{conect}) in the Appendix \ref{vdwhc}.

The   hydrodynamic equations (\ref{fin2})    are equivalent  to  those
proposed by   Felderhof \cite{Felderhof70}. However,   in  Felderhof's
theory,   dissipation  was   neglected   and,   consistently,  thermal
fluctuations where not  considered.  Thus, equations (\ref{fin2}) with
(\ref{ta}) and (\ref{qv}) can  be taken as the natural  generalization
of Felderhof theory.

It is not obvious  at a first glance that   the total momentum of  the
system is conserved by  the term $c\rho_{\bf r}\nabla\nabla^2\rho_{\bf
r}$ in (\ref{fin2}).  That this is the case can be seen by noting that
this term can be cast into the form $-\nabla\!\cdot\!{\bf P}_{\bf r}$,
and  the momentum equation has the  form of a   balance equation.  The
``surface  tension'' contribution to  the stress  tensor ${\bf P}_{\bf
r}$ has the form \cite{note1}

\begin{equation}
{\bf P}_{\bf r} = 
2 c \left[\frac{1}{6}\nabla \rho_{\bf r}\nabla \rho_{\bf r}
+\frac{1}{12}(\nabla \rho_{\bf r})^2{\bf 1}
-\frac{1}{3}\rho_{\bf r}\nabla\nabla \rho_{\bf r}
-\frac{1}{6}\rho_{\bf r}\nabla^2 \rho_{\bf r}{\bf 1}\right].
\label{stress}
\end{equation}
This contribution to the stress tensor is identical to the one derived
from molecular considerations  for the {\em equilibrium} stress tensor
\cite{Irwing50}  under the  gradient approximation\cite{Davis96}.   We
observe, therefore, that   the theory that  is  being presented relies
also on  the hypothesis of {\em  local equilibrium}, in  much the same
way  as  in  the  usual    continuum  hydrodynamics  approach.   
 
\section{Summary and discussion}
\label{discussion}
We  have computed the entropy  functional  for a simple fluid starting
from microscopic principles under the classic van der Waals assumption
that the  molecular potential  has  two well-separated  ranges, a hard
core repulsive  part plus a very long  ranged attractive  tail.  Under
the further assumption that the hydrodynamic fields vary slowly on the
total range of the  potential it is possible to  write a local entropy
functional  that  depends  on  the  density  gradients.   We have also
presented the connection between this entropy functional and the local
gradient free energy  functional which is used  in the  context of the
study of equilibrium   of  liquid-vapor interfaces.  Even  though  the
assumption  of  the unrealistic molecular   potential seems  to be too
restrictive,  the equilibrium gradient theory  is highly successful in
describing the equilibrium  properties of these interfaces,  and gives
confidence on the dynamical approach taken in this paper.

The  entropy  functional describes  the  equilibrium properties of the
system but  it is also  one of the basic  inputs  for constructing the
non-equilibrium evolution  equations   for  the hydrodynamic    fields
through the GENERIC  formalism.  By making  use of all  the well-known
information about the $L$ and $M$ matrices  for the hydrodynamics of a
single phase fluid  flow, we can  construct the hydrodynamic equations
for a van der Waals fluid in a rather simple way.

The  hydrodynamic equations proposed predict   unstable forces if  the
thermodynamic state is in the positive slope part of the van der Waals
pressure diagram. When   this happens, the fluid  spontaneously adopts
one   of   the two  possible  densities   corresponding  to liquid and
vapor.  Density   variations  appear  in   the  short  length  scales
associated to the interfaces of bubbles or drops.   It is precisely in
these interfacial regions where the  last term in Eqn. (\ref{mom1}) is
important. This term  generates forces in the  fluid associated to the
surface  tension of  the  interface. Actually,  the interfaces  try to
adopt a spherical shape.   The interface is  a  diffuse object  with a
finite width.  Note  that no boundary  conditions are required on  the
interface of a bubble or a drop.  Actually, there  is no need to known
a priori the   location  of the ``interface''    in order  to have   a
well-posed hydrodynamic problem.

The set  of  equations  (\ref{fin}) with the   gradient  approximation
(\ref{appot}) and the set of equations (\ref{fin2}) are mathematically
equivalent.  We  regard as    the  main benefit  of    the microscopic
derivation presented in  this paper the clear physical  interpretation
of the variables (either $\epsilon^i_{\bf  r}$ or $u_{\bf r}$) and the
correct equations of state  (either $P^{\rm hc}$  or $P^{\rm vdW}$) to
be used in the equations.  This issue is not always clear in the usual
presentations of this subject based on phenomenological arguments that
simply   add gradient terms  either    in the  energy or  the  entropy
\cite{Anderson98}.

 From the  computational   point of view   of numerically  solving the
hydrodynamic   equations   obtained, we  observe  that  a  theory like
Kawasaki's   in which the  mean   field long  range  forces appear  in
integral  form, Eqns.   (\ref{fin}), might  represent an advantage  in
front  of a theory like  Felderhof's  in which  these forces appear in
differential form through  third   derivatives of the density   field,
Eqns.   (\ref{fin2}).  In   those  regions where   the  interfaces are
located, the density field changes strongly which means that the third
derivatives present a rather   complex structure in very short  length
scales.  In  order  to get  stable and  convergent results, very  fine
grids are required.  These problems do not arise in the discretization
of the integral form of the long range forces.

The theory  derived here  connects   with the well-known  phase  field
theories  that  deal   with  systems  that  present  complex  boundary
conditions like   those occurring  in   melting and  dendritic  growth
\cite{Karma98} or immiscible  fluids \cite{Folch99}. In these systems,
the partial differential equations governing the evolution of the
relevant variables require the  formulation of boundary conditions  on
moving surfaces. The motion  of the surface is,  in turn, coupled in a
complex way with the dynamics of the  relevant variables. The approach
taken by the phase field method  is to include the boundary conditions
into the equations of motion  through the introduction of an auxiliary
phase field that locates in a somewhat diffuse way the position of the
interface.  A further dynamics is generated  for the phase field which
is coupled with the dynamics of the rest of fields in the system. This
dynamics is obtained from a  phenomenological formulation of a  ``free
energy''.  The theory we have derived is then  in the class of a phase
field theory for bubbles and drops. In our case, the phase field has a
definite  physical meaning in   terms  of the   density field,  and no
phenomenological free-energy needs to be introduced ad hoc.

P.E.  would  like to thank   useful discussions with  H.C. \"Ottinger,
P. Tarazona, J. Cuesta, and R. Delgado.  This work has been partially
supported by DGYCIT PB97-0077.

\section{Appendix: Microscopic derivation of the entropy functional}
\label{ap1}
In   this  section,  we   compute from   first principles  the   joint
probability  that  an  extended simple   fluid   in equilibrium  has a
particular realization of the   mass,  momentum and  internal   energy
density  fields. This generalizes  van   Kampen derivation which  only
considered the mass density field \cite{vanKampen64}.

The isolated fluid is assumed to  be described microscopically by $z$,
the set of positions ${\bf q}_i$ and momenta ${\bf p}_i$ of the center
of mass  of   its  $N_0$  constituent   molecules (no   rotational  or
vibrational  degrees of freedom  are considered,  for simplicity). The
Hamiltonian of the system is given by

\begin{equation}
H(z) = \sum_i^{N_0} \frac{{\bf p}_i^2}{2m}+\frac{1}{2}\sum_{ij}\phi_{ij},
\label{hamiltonian}
\end{equation}
where $m$  is the mass of a  molecule and $\phi_{ij}$ is the pair-wise
potential function between molecules  $i$  and $j$, assumed to  depend
only on its separation  $r_{ij}$. The total momentum  of the system is
given by
\begin{equation}
{\bf P}(z) = \sum_i^{N_0} {\bf p}_i.
\label{momentum}
\end{equation}
The  microscopic dynamics  is such  that $H(z)$ and  ${\bf  P}(z)$ are
dynamical invariants. Obviously,  the total number $N_0$  of particles
(or the total mass) is also a dynamical invariant.

On a  mesoscopic scale, the  space containing the fluid is partitioned
in $M$ cells. The state of the system at that  scale is described with
the      mesoscopic variables    $N_\mu,{\bf     P}_\mu,\epsilon_\mu$,
$\mu=1,\ldots,M$,  where $N_\mu$  is  the number of  particles in cell
$\mu$, ${\bf P}_\mu$ is   the momentum of  the center  of mass of  the
particles of cell $\mu$ and  $\epsilon_\mu$ is the energy with respect
to  the center of  mass of the particles  in cell  $\mu$.  In order to
relate    the mesoscopic variables    with   the microscopic ones,  we
introduce  the characteristic  function  $\chi_\mu({\bf  r})$  of cell
$\mu$ which takes the value 1 if ${\bf r}$ is within  cell $\mu$ and 0
otherwise. For  example,  the  space can be  partitioned  into Voronoi
cells  whose   characteristic function  is $\chi_{\mu}({\bf  r})\equiv
\prod_{\nu  \neq  \mu}^M\theta(|{\bf  r}-{\bf  R}_\nu| -|{\bf  r}-{\bf
R}_\mu|)$ where $\theta(x)$ is the  Heaviside step function and  ${\bf
R}_\mu$ is the center of the Voronoi cell $\mu$. Obviously,

\begin{equation}
\sum^M_\mu \chi_\mu({\bf r})=1,
\label{closed}
\end{equation}
because the cells cover all space without overlapping. The volume of
the cell is given by
\begin{equation}
{\cal V}_\mu= \int_{V_0}d^3{\bf r}\chi_{\mu}({\bf r}),
\end{equation}
where $V_0$ is the total volume of the system. 
The mesoscopic variables $N_\mu$, ${\bf P}_\mu$ and $\epsilon_\mu$
can now be written as functions of the microstate $z$, this is

\begin{eqnarray}
N_\mu(z) &=& \sum_i\chi_\mu({\bf q}_i),
\nonumber\\
{\bf P}_\mu(z) &=& \sum_i{\bf p}_i\chi_\mu({\bf q}_i),
\nonumber\\
\epsilon_\mu(z) &=& \sum_i\left(\frac{1}{2m}
\left({\bf p}_i-\frac{{\bf P}_\mu}{N_\mu}\right)^2
+\phi_i\right)\chi_\mu({\bf q}_i),
\label{Az}
\end{eqnarray}
where we have introduced the potential energy of particle $i$ through

\begin{equation}
\phi_i = \frac{1}{2}\sum_j \phi_{ij}.
\label{phii}
\end{equation}
Note that  we   use  Latin indices  to   refer  to  variables  at  the
microscopic level and Greek indices to refer to mesoscopic variables.

The dynamical  invariants can  be written in  terms of  the mesoscopic
variables by using Eqn. (\ref{closed}), this is
\begin{eqnarray}
H(z)&=&\sum_\mu\frac{{\bf P}_\mu^2(z)}{2M_\mu(z)}+\epsilon_\mu(z),
\nonumber\\
{\bf P}(z)&=&\sum_\mu{\bf P}_\mu(z),
\label{HPmes}
\end{eqnarray}
where $M_\mu=mN_\mu$ is the mass of cell $\mu$.

We will assume that the system is at equilibrium, which means that the
effective  probability density   $\rho^{\rm   eq}(z)$ of  finding    a
particular value of  $z$ is a function of  the dynamical invariants of
the system only.  If we assume  that  these invariants are known  with
precision,  the equilibrium  distribution   function is given  by  the
molecular ensemble

\begin{equation}
\rho^{\rm eq}(z) 
=\frac{1}{N_0!\Omega_0}\delta({\bf P}(z)-{\bf P}_0)
\delta(H(z)-E_0),
\label{mol}
\end{equation}
where ${\bf P}_0$ is the  total momentum of  the system and $E_0$  the
total energy. The factor $N_0!$ is quantum  mechanically in origin (it
comes from  the undistinguishability of  the particles) and solves the
Gibbs paradox   (and  makes the   macroscopic entropy  a  first  order
function of energy and number  of particles). The normalization factor
$\Omega_0$ is given by

\begin{equation}
\Omega_0 = \Omega(U_0,N_0,V_0)=\int \frac{dz}{N_0!}
\delta({\bf P}(z)-{\bf P}_0)
\delta(H(z)-E_0).
\label{om0}
\end{equation}
Here, $U_0=E_0-P^2_0/2M_0$ is the  internal energy of the total system
and the total  volume $V_0$  appears parametrically  as  the region of
integration  of positions.   The  {\em macroscopic} entropy for   this
system is defined as
\begin{equation}
S^{\rm mac}(U,N,V) = k_B \ln \Omega_0(U,N,V).
\label{macS}
\end{equation}
The intensive parameters  are   defined  as the derivatives  of    the
macroscopic entropy, this is,
\begin{eqnarray}
\frac{1}{T}&=&\frac{\partial S^{\rm mac}}{\partial U},
\nonumber\\
\frac{\mu}{T}&=&\frac{\partial S^{\rm mac}}{\partial N},
\nonumber\\
-\frac{P}{T}&=&\frac{\partial S^{\rm mac}}{\partial V},
\label{intensive2}
\end{eqnarray}
where $T$  is the temperature, $\mu$ the  chemical  potential, and $P$
the pressure.

\subsection{Calculation of $P[N,P,\epsilon]$}

The probability that a set of functions $X(z)$ take particular values
$x$ when the system is at equilibrium is given by

\begin{equation}
P(x) = \int dz \delta(X(z)-x)\rho^{\rm eq}(z),
\label{palfa}
\end{equation}
where $\rho^{\rm eq}(z)$ is the equilibrium ensemble.  We now consider
the functions $X(z)$ that  appear in Eqn.  (\ref{palfa}) to be the set
of mesoscopic variables (\ref{Az}). In this  way, the probability that
the system  adopts    a   particular   set of   values     $N_\mu,{\bf
P}_\mu,{\epsilon}_\mu$ for each cell is given by

\begin{eqnarray}
P[N,{\bf P},{\epsilon}]
&=&
\int dz \rho^{\rm eq}(z)\prod_\mu^M
\chi\left(N_\mu(z)-N_\mu\right)
\nonumber\\
&\times&
\delta\left({\bf P}_\mu(z)-{\bf P}_\mu\right)
\delta\left(\epsilon_\mu(z)-{\epsilon}_\mu\right).
\label{C1}
\end{eqnarray}
In this expression,   we have introduced  the  characteristic function
$\chi\left(N_\mu(z)-N_\mu\right)$  that    takes the value    $1$ (not
infinity) for the  region of phase space in  which the  microstate $z$
produces exactly the value $N_\mu$ for the number of particles in cell
$\mu$. The probability (\ref{C1}) is normalized according to
\begin{equation}
\int d{\bf P}_1\ldots d{\bf P}_M
\int d\epsilon_1\ldots d\epsilon_M
{\sum_{N_1,\ldots,N_M}^{N_0}}
P[N,{\bf P},{\epsilon}]= 1,
\label{normp}
\end{equation}
where the numbers $N_\mu$ are  subject to $\sum_\mu  N_\mu = N_0$.  By
using the equilibrium ensemble (\ref{mol})  and Eqns. (\ref{HPmes}) we
obtain
\begin{eqnarray}
P[N,{\bf P},{\epsilon}]
&=&
\frac{1}{\Omega_0}
\delta\left(\sum_\mu^M{\bf P}_\mu-{\bf P}_0\right)
\nonumber\\
&\times&
\delta\left(\sum_\mu^M\left(\frac{{\bf P}^2_\mu}{2M_\mu}+\epsilon_\mu\right)
-E_0\right)
\nonumber\\
&\times&
\int \frac{dz}{N_0!}
\prod_\mu^M\chi(N_\mu(z)-N_\mu)\delta({\bf P}_\mu(z)-{\bf P}_\mu)
\nonumber\\
&\times&
\delta(\epsilon_\mu(z)-\epsilon_\mu).
\label{C2}
\end{eqnarray}
The term $\prod_\mu^M\chi(N_\mu(z)-N_\mu)$ within the integral is zero
unless the  microstate  $z$  is  such that   there are  exactly  $N_1$
particles  in the  first cell, $N_2$  in the   second, etc.  There are
$N_0!/N_1! \cdots N_M!$ ways of having the $N_0$ particles distributed
among   the  $M$   cells    with  the  prescribed   numbers in    each
cell. Therefore, we can write
\begin{eqnarray}
&&\int \frac{dz}{{N_0}!}
\prod_\mu^M\chi(N_\mu(z)-N_\mu)\delta({\bf P}_\mu(z)-{\bf P}_\mu)
\delta(\epsilon_\mu(z)-\epsilon_\mu)
\nonumber\\
&&=
\frac{1}{N_1!\cdots N_M!}
\int_{{\cal V}_1} \underbrace{d{\bf q}\ldots d{\bf q}}_{N_1} \ldots
\int_{{\cal V}_M} \underbrace{d{\bf q}\ldots d{\bf q}}_{N_M}
\nonumber\\
&&
\times \int d{\bf p}_1\ldots d{\bf p}_M\prod_\mu^M
\delta\left(\sum_{i_\mu}^{N_\mu}{\bf p}_{i_\mu}-{\bf P}_\mu\right)
\nonumber\\
&&\times
\delta\left(\sum_{i_\mu}^{N_\mu}\left(\frac{1}{2m}
\left({\bf p}_{i_\mu}-\frac{{\bf P}_\mu}{N_\mu}\right)^2
+\phi_{i_\mu}\right)-\epsilon_\mu\right)
\nonumber\\
&&
=
\frac{1}{N_1!\cdots N_M!}
\int_{{\cal V}_1} \underbrace{d{\bf q}_1\ldots d{\bf q}_{N_1}}_{N_1}
\cdots\int_{{\cal V}_M} \underbrace{d{\bf q}\ldots d{\bf q}_{M}}_{N_M}
\nonumber\\
&&
\times\prod_\mu^M \Phi({\bf 0},\epsilon_\mu-\sum_{i_\mu}^{N_\mu}\phi_{i_\mu},
N_\mu),
\label{C3}
\end{eqnarray}
where the function  $\Phi({\bf P},E,N)$ is  introduced and computed in
the Appendix \ref{apmol}. The   numbers $N_1,\ldots,N_M$ are subject  to $\sum_\mu
N_\mu=N_0$ and, therefore, we can extend the above result to arbitrary
$N_1,\ldots,N_M$  provided that we multiply  Eqn.  (\ref{C3}) with the
factor $\chi(\sum_\mu N_\mu  -N_0)$, which  takes  the value 1  if its
argument is zero.

Let us write now the potential energy of cell $\mu$ as follows,

\begin{equation}
\sum_{i_\mu}^{N_\mu} \phi_{i_\mu}
=
\frac{1}{2}\sum_{i_\mu}^{N_\mu}\sum_{j_\mu}^{N_\mu}
\phi_{i_\mu j_\mu}
+
\frac{1}{2}\sum_\nu'
\sum_{i_\mu}^{N_\mu}\sum_{j_\nu}^{N_\nu}
\phi_{i_\mu j_\nu}.
\label{C4}
\end{equation}
The first term is  the potential energy due  to the particles  of cell
$\mu$ and it is a function solely of the  coordinates of the particles
that are in cell $\mu$.  The second  term, the potential energy due to
the interaction of particles  in different cells,  involve coordinates
of particles in cells $\nu$ different from $\mu$. It is this last term
which hinders  the decoupling  of the integrals  in  positions in Eqn.
(\ref{C3}). For this reason, it is necessary to make approximations to
this last term.   In what follows,  we will approximate the  potential
energy due to different cells by a mean field constant (independent of
coordinates of particles), this is,

\begin{equation}
\frac{1}{2}\sum_\nu'
\sum_{i_\mu}^{N_\mu}\sum_{j_\nu}^{N_\nu}
\phi_{i_\mu j_\nu}
\approx\sum_\nu'\frac{1}{2}N_\mu N_\nu C_{\mu\nu}.
\label{C5}
\end{equation}
This approximation is fully justified when the potential $\phi(r)$ has
two typical ranges,  a short length scale $r_s$  much smaller than the
typical  size of Voronoi cells  and  a long  length  scale $r_l$  much
larger  than  the size of the  cells.  In this case,  we  can separate
$\phi(r)=\phi^s(r)+\phi^l(r)$ and write
\begin{equation}
\frac{1}{2}\sum_{i_\mu}^{N_\mu}\sum_{j_\nu}^{N_\nu}
\phi_{i_\mu j_\nu}
\approx \frac{1}{2}
\sum_{i_\mu}^{N_\mu}\sum_{j_\nu}^{N_\nu}
\phi^l_{i_\mu j_\nu},
\label{C7}
\end{equation}
because  the   short range part  is  negligible  for   different cells
$\mu\neq\nu$. Also, because  the range $r_l$ is  much larger than  the
size of a cell, we  can further approximate  Eqn. (\ref{C7}) as if all
particles  in each cell  where located at  the center of the cell and,
therefore, we can write

\begin{equation}
\frac{1}{2}
\sum_{i_\mu}^{N_\mu}\sum_{j_\nu}^{N_\nu}
\phi^l_{i_\mu j_\nu}\approx 
\frac{1}{2}\phi^l(R_{\mu\nu})N_\mu N_\nu,
\label{C8}
\end{equation}
and  the potential energy due to  different cells has the structure of
Eqn. (\ref{C5}).

Under the mean field approximation  (\ref{C5}), the different position
integrals  in  (\ref{C3})  decouple, so  that Eqn.   (\ref{C3}) can be
written as
\begin{equation}
\prod_\mu^M\left[ \frac{1}{N_\mu!}
\int_{{\cal V}_\mu}
d{\bf q}_{1_\mu}\ldots d{\bf q}_{N_\mu}
\Phi({\bf 0},\epsilon_\mu-\phi_\mu, N_\mu)\right],
\label{C9}
\end{equation}
where
\begin{eqnarray}
\phi_\mu&=&\phi_\mu({\bf q}_{1_\mu},\ldots,{\bf q}_{N_\mu})
=\frac{1}{2}\sum_{i_\mu}^{N_\mu}\sum_{j_\mu}^{N_\mu}\phi_{i_\mu j_\mu}
+\overline{\phi}_\mu
\nonumber\\
\overline{\phi}_\mu &=&\frac{1}{2}\sum_\nu'\phi^l(R_{\mu\nu})N_{\mu}N_{\nu},
\label{C10b}
\end{eqnarray}
where $\overline{\phi}_\mu$ is  the  mean field inter-cell   potential
energy of cell $\mu$.

We can write Eqn. (\ref{C9}) as
\begin{equation}
\prod_\mu^M \Omega^{\rm hc}(\epsilon_\mu-\overline{\phi}_\mu,N_\mu,{\cal V}_\mu),
\label{C9b}
\end{equation}
where $\Omega^{\rm hc}(U,N,V)$ is defined  in Eqn. (\ref{om0}).  Note  that the
Hamiltonian that should appear in Eqn. (\ref{om0}) is that of a system
of particles interacting through the short range part of the potential
$\phi^s(r)$. Finally, Eqn. (\ref{C2}) can be written as

\begin{eqnarray}
P[N,{\bf P},{\epsilon}]
&=&
\frac{1}{\Omega_0}\chi\left(\sum^M_\mu N_\mu - N_0\right)
\delta\left(\sum_\mu^M{\bf P}_\mu-{\bf P}_0\right)
\nonumber\\
&\times&
\delta\left(\sum_\mu^M
\left(\frac{{\bf P}^2_\mu}{2M_\mu}+\epsilon_\mu\right)-E_0\right)
\nonumber\\
&\times&
\exp \left\{S[N,{\epsilon}]/k_B\right\},
\label{C14}
\end{eqnarray}
where the entropy functional at the mesoscopic level is defined by
\begin{equation}
S[N,{\epsilon}]
=\sum_\mu S^{\rm hc}(\epsilon_\mu
-\overline{\phi}_\mu,N_\mu,{\cal V}_\mu),
\label{mesentap}
\end{equation}
where we have used Eqn. (\ref{macS}).  The mesoscopic entropy is given
in terms of  the sum of the  {\em macroscopic} entropies of each cell,
as if  they were  isolated systems  in a volume  ${\cal V}_\mu$,  with
number   of      particles    $N_\mu$      and   with      an   energy
$\epsilon_\mu-\overline{\phi}_\mu$.     The        internal     energy
$\epsilon_\mu$ in  (\ref{Az})  includes in its  microscopic definition
the potential energy of  particles  that  interact with particles   of
neighboring cells,  that  is, $\epsilon_\mu$   includes the inter-cell
energy. In this way, $\epsilon_\mu-\overline{\phi}_\mu$ represents the
purely internal energy  due to the particles within  the cell (in mean
field).  Even though we have made progress  in writing the probability
for the mesoscopic variables, Eqn.   (\ref{C14}), the entropy of  each
cell $S^{\rm hc}(\epsilon_\mu,{\cal V}_\mu,N_\mu)$ is still an unknown
quantity.  We remark that this entropy is the macroscopic entropy of a
system   of molecules interacting   with the short  range  part of the
potential. In applications of the model, this entropy will be given by
simple  models   (see  Appendix  \ref{vdwhc})   or  by very   accurate
expressions      like the    Carnahan-Starling    equation   of  state
\cite{Davis96}.

Apparently, Eqn.  (\ref{mesentap})  simply says that  ``the entropy is
additive''.     However, this   sentence   is  imprecise:  The entropy
$S[N,\epsilon]$ in the lhs  of  Eqn. (\ref{mesentap}) is a   different
object from the entropy $S^{\rm hc}(\epsilon_\mu,N_\mu,{\cal V}_\mu)$
appearing in the rhs. They depend on a  different number of variables.
Therefore, one should rather say that ``the  mesoscopic entropy at the
level of hydrodynamic variables is the sum of the entropy at the level
of dynamical invariants (the macroscopic  entropy) of each cell as  if
they  were isolated''. 

The macroscopic entropy is a first order function of its variables and
therefore,
\begin{equation}
S[N,{\epsilon}]
=\sum_\mu{\cal V}_\mu S^{\rm hc}
\left(\frac{\epsilon_\mu}{{\cal V}_\mu}
-\frac{\overline{\phi}_\mu}{{\cal V}_\mu},\frac{N_\mu}{{\cal V}_\mu},1\right)
\label{mesent2}
\end{equation}
which admits the continuum {\em notation}
\begin{equation}
S[n_{\bf r},{\epsilon}_{\bf r}]
=\int d{\bf r} s^{\rm hc}
(\epsilon_{\bf r}-\overline{\phi}_{\bf r},n_{\bf r}),
\label{mesent3}
\end{equation}
where we  have   introduced the continuum   version of   the effective
interaction energy between   molecules  in different points   of space
(cells),

\begin{eqnarray}
{\bf R}_\mu &\rightarrow& {\bf r},
\nonumber\\
\sum_\mu{\cal V}_\mu&\rightarrow&\int d{\bf r},
\nonumber\\
\frac{\epsilon_\mu}{{\cal V}_\mu}&\rightarrow&\epsilon_{\bf r},
\nonumber\\
\frac{N_\mu}{{\cal V}_\mu}&\rightarrow&n_{\bf r},
\nonumber\\
\frac{\overline{\phi}_\mu}{{\cal V}_\mu}&\rightarrow&\overline{\phi}_{\bf r}.
\label{densities}
\end{eqnarray}
Note that in this continuum notation we can write the dynamical
invariants (\ref{HPmes}) of the system as
\begin{eqnarray}
E&=&\int d{\bf r}\left(\frac{1}{2}
\frac{{\bf g}_{\bf r}}{mn_{\bf r}}+\epsilon_{\bf r}\right),
\nonumber\\
P&=& \int d{\bf r}{\bf g}_{\bf r},
\label{incon}
\end{eqnarray}
where ${\bf   P}_\mu/{\cal V}_\mu\rightarrow {\bf   g}_{\bf  r}$ is the
momentum density field.  We stress that we regard Eqn. (\ref{mesent3})
not as  an strict mathematical limit in  which the volume of the cells
tend to zero, but rather as a convenient notational tool.

In spite of its  apparent form, the mesoscopic entropy (\ref{mesent3})
is {\em not local} in  space. This is,  it is not given  by a sum of a
function  of the variables   at  a given point.   This is  due to  the
presence of the term $\overline{\phi}_{\bf r}$ which can be written in
continuous notation as

\begin{equation}
\overline{\phi}_{\bf r}= \int d{\bf r}' \phi^l(|{\bf r}-{\bf r}'|)
n_{\bf r}n_{{\bf r}'}.
\end{equation}

We have implicitly associated the potential energy of interaction with
other cells $\overline{\phi}_{\bf  r}$ with  the phenomena of  surface
tension.  In order  to  make this   connection more  explicit, in  the
following   subsections  we    consider  the    marginal   probability
distribution functions   and   will make  contact  with   the  Density
Functional Theory.

\subsection{The marginal distribution $P[N,\epsilon]$}
By integrating the  distribution function $P[N,{\bf P},\epsilon]$ over
momenta we   will have  the   probability of  a  realization  of   the
``fields'' $N,\epsilon$  irrespective of the values  of the momenta in
each cell, this is

\begin{eqnarray}
P[N,\epsilon] &=&\exp \left\{S[N,{\epsilon}]/k_B\right\}
\frac{1}{\Omega_0}\chi\left(\sum^M_\mu N_\mu - N_0\right)
\nonumber\\
&\times&
\int d{\bf P}_1\ldots d{\bf P}_M
\delta\left(\sum_\mu^M{\bf P}_\mu-{\bf P}_0\right)
\nonumber\\
&\times&
\delta\left(\sum_\mu^M
\left(\frac{{\bf P}^2_\mu}{2M_\mu}+\epsilon_\mu\right)-E_0\right)
\nonumber\\
&=&\exp \left\{S[N,{\epsilon}]/k_B\right\}
\frac{1}{\Omega_0}\chi\left(\sum^M_\mu N_\mu - N_0\right)
\nonumber\\
&\times&
\prod_\mu^M(2mN_\mu)^{D/2}\frac{\omega_{D(M-1)}}{2}
\left[U_0-\sum_\mu\epsilon_\mu\right]^{\frac{D(M-1)}{2}-1},
\nonumber\\
\label{pne}
\end{eqnarray}
where we have used once more Eqn. (\ref{11}) in the appendix.

The            most            probable                    realization
$N_1^*,\ldots,N_M^*,\epsilon_1^*,\ldots,\epsilon_M^*$ of the fields is
the one which maximizes the functional

\begin{eqnarray}
&&k_B^{-1}S[N,\epsilon]
+\left(\frac{D(M-1)}{2}-1\right)\log\left(U_0-\sum_\mu\epsilon_\mu\right)
\nonumber\\
&&
+\frac{D}{2}\sum_\mu \log (2m N_\mu)+\beta\lambda\sum_\mu N_\mu,
\label{funct}
\end{eqnarray}
where we have introduce the Lagrange multiplier $\beta \lambda$ that
takes into account the restriction $\sum_\mu N_\mu=N_0$
The maximum $N^*,\epsilon^*$ of the functional (\ref{funct}) is the
solution of the
following set of equations
\begin{eqnarray}
\frac{\partial S}{\partial N_\mu}
[N^*,\epsilon^*]&=&k_B\beta\lambda,
\nonumber\\
\frac{\partial S}{\partial \epsilon_\mu}[N^*,\epsilon^*]
&=&\frac{(\frac{D(M-1)}{2}-1) k_B}
{U_0-\sum_\nu\epsilon_\nu^*},
\nonumber\\
\sum_\mu N_\nu^*&=& N_0.
\label{inte}
\end{eqnarray}
We have neglected a term of  order $N_\mu^{-1}$ in the first equation,
which  is reasonable  if typically there  are  many particles  in each
cell. Eqns. (\ref{inte}) is a set of $2M+1$ equations for the unknowns
$N_\mu^*,\epsilon_\mu^*,\lambda$.     Note   that    the      solution
$N_\mu^*,\epsilon_\mu^*,\beta\lambda$    depends   parametrically   on
$N_0,U_0$.

We find now a convenient  approximation to Eqn. (\ref{pne}) by  noting
that this probability is expected to  be highly peaked around the most
probable state.  Therefore,  for those values  of the field $\epsilon$
for which $P[N,\epsilon]$ is appreciably different from zero (that is,
around $\epsilon^*$), we can approximate

\begin{eqnarray}
&&\left[U_0-\sum_\mu\epsilon_\mu\right]^P
= \left[U_0-\sum_\mu\epsilon_\mu^*\right]^P
\nonumber\\
&\times&\left[1
+\frac{\beta}{P}\sum_\mu(\epsilon_\mu^*-\epsilon_\mu)\right]^P
\nonumber\\
&\approx&
\left[U_0-\sum_\mu\epsilon_\mu^*\right]^P
\exp\{\beta\sum_\mu \left(\epsilon_\mu^*-\epsilon_\mu\right)\}
\nonumber\\
&=&{\rm ctn.} \exp \{-\beta\sum_\mu\epsilon_\mu\},
\label{approx}
\end{eqnarray}
where $P=D(M-1)/2-1$ is a very large number and we have introduced
\begin{equation}
\beta=\frac{D(M-1)/2-1}{U_0-\sum_\nu\epsilon_\nu^*}.
\label{beta}
\end{equation}
Note that $\beta$ is proportional to  the inverse of the most probable
value of the kinetic energy per cell.

Finally, we can write Eqn. (\ref{pne}) as
\begin{eqnarray}
P[N,\epsilon] 
&=&\frac{1}{\Omega'_0}
\exp \left\{S[N,{\epsilon}]/k_B-\beta\sum_\mu\epsilon_\mu\right\}
\nonumber\\
&\times&\chi\left(\sum_\mu N_\mu - N_0\right),
\label{pne2}
\end{eqnarray}
where $\Omega'_0$ is  the corresponding  normalization function.    We
see, therefore, that by integrating the momenta the ``microcanonical''
form Eqn. (\ref{C14}) becomes the ``canonical'' form (\ref{pne2}).

\subsection{The marginal distribution $P[N]$}

There are two different routes to compute the probability of a certain
distribution $N_1,\ldots,N_M$ of the particles  in the cells. We could
simply start the calculation from Eqn. (\ref{C1}) without the momentum
and energy  conserving delta functions.  This route is essentially the
one taken  by van  Kampen  \cite{vanKampen64}. The  alternative is  to
integrate out the energy field in  Eqn. (\ref{pne2}). The same results
are obtained  in both approaches  and  we illustrate  here this second
one, this is,

\begin{eqnarray}
P[N]&=&\int d\epsilon_1 \ldots d\epsilon_M P[N,\epsilon]
\nonumber\\
&=&\frac{1}{Z_0}
\exp \left\{-\beta\sum_\mu {
F^{\rm hc}(\beta,N_\mu,{\cal V}_\mu)-\beta\overline{\phi}_\mu}
\right\}
\nonumber\\
&\times&\chi\left(\sum^M_\mu N_\mu - N_0\right),
\label{pnap}
\end{eqnarray}
where we have used the form (\ref{mesentap}) for the entropy function,
we      have       changed       variables       to    $\epsilon_\mu'=
\epsilon_\mu-\overline{\phi}_\mu$ and have introduced the  macroscopic
free  energy $F^{\rm hc}(\beta,N,V)$ of  a  system of $N$ particles at
inverse   temperature $\beta$ in a  volume  V, and  interacting with a
short range potential. In deriving  Eqn. (\ref{pnap}) we have made use
of the following identity
\begin{equation}
\int_0^\infty \exp\{S(U)/k_B-\beta U\} dU =\exp\{-\beta F(\beta)\}.
\label{rig}
\end{equation}
This expression is the statistical mechanics link between the
macroscopic entropy and the free energy and can be proved by computing the Laplace
transform of $\Omega_0$ in Eqn. (\ref{om0}). One has

\begin{eqnarray}
&&\int_0^\infty \Omega_0(U,N,V)\exp-\{\beta U\}dU 
\nonumber\\
&&=
\int dz \delta(\sum_i{\bf p}_i)\exp-\{\beta H(z)\}
= Z(\beta,N-1,V),
\label{part}
\end{eqnarray}
where   the partition  function  $Z(\beta)$   is defined  through this
equation. The presence or  absence of the momentum conservation  delta
function is irrelevant when the number of particles is large so we may
very well drop it.  The free energy $F(\beta)$ is defined as
\begin{equation}
\beta F(\beta)=-\ln Z(\beta),
\label{freee}
\end{equation}
The usual thermodynamic link between entropy and free energy,

\begin{equation}
F(T,N,V)=  U - TS(U,N,V),
\label{fut}
\end{equation}
can be obtained under the assumption that the integrand in (\ref{rig})
is highly peaked.

\section{Appendix: Molecular ensemble}
\label{apmol}
In this appendix we compute explicitly the following integral

\begin{eqnarray}
\Phi({\bf P}_0,E_0,M)&=&\int d^{DM}{\bf p}
\,\,\delta\!\left(\sum_i^M \frac{{\bf p}_i^2}{2m_i}-E_0\right)
\nonumber\\
&\times&
\delta^D\!\left(\sum_i^M{\bf p}_i-{\bf P}_0\right),
\label{4}
\end{eqnarray}
which  appears repeatedly  when  computing  molecular averages.   
By a simple change of variables we obtain

\begin{eqnarray}
\Phi({\bf P}_0,E_0,M)&=&
\prod_i^M(2m_i)^{D/2}\int d^{DM}{\bf p}
\,\,\delta\!\left(\sum_i^M {\bf p}_i^2-E_0\right)
\nonumber\\
&\times&
\delta^D\!\left(\sum_i^M(2m_i)^{1/2}
{\bf p}_i-{\bf P}_0\right),
\label{4bis}
\end{eqnarray}
 The equation $\sum_i^M(2m_i)^{1/2}{\bf p}_i={\bf P}_0$ are actually D
equations  (one  for each  component of the  momentum)  which define D
planes  in $R^{DM}$. The  integral  in  (\ref{4}) is  actually  over a
submanifold which is the intersection of the D planes with the surface
of a $DM$ dimensional sphere of  radius $E_0^{1/2}$. This intersection
will be also a sphere, which will be now of smaller radius and also of
smaller dimension, $D(M-1)$.

In order to compute (\ref{4}), we change to the following notation

\begin{eqnarray}
\mbox{\boldmath ${\cal P}$}
&=&(p_1^x,\ldots,p_M^x,p_1^y,\ldots,p_M^y,p_1^z,\ldots,p_M^z)
\nonumber\\
\mbox{\boldmath ${\cal C}$}_x &=&
((2m_1)^{1/2},\ldots,(2m_M)^{1/2},0,\ldots,0,0,\ldots,0)
\nonumber\\
\mbox{\boldmath ${\cal C}$}_y &=&
(0,\ldots,0,(2m_1)^{1/2},\ldots,(2m_M)^{1/2},0,\ldots,0)
\nonumber\\
\mbox{\boldmath ${\cal C}$}_z &=&
(0,\ldots,0,0,\ldots,0,(2m_1)^{1/2},\ldots,(2m_M)^{1/2}).
\label{defis}
\end{eqnarray}
Note that these vectors satisfy
 $\mbox{\boldmath ${\cal C}$}_x\!\cdot\!\mbox{\boldmath ${\cal C}$}_y=0$
 $\mbox{\boldmath ${\cal C}$}_y\!\cdot\!\mbox{\boldmath ${\cal C}$}_z=0$
 $\mbox{\boldmath ${\cal C}$}_z\!\cdot\!\mbox{\boldmath ${\cal C}$}_x=0$. 
With these vectors so defined, Eqn. (\ref{4}) becomes
\begin{eqnarray}
\Phi({\bf P}_0,E_0,M)
&=&\int d^{DM}\mbox{\boldmath ${\cal P}$}
\delta\!\left(\mbox{\boldmath ${\cal P}$}^2-E_0\right)
\,\,\delta\!\left(
\mbox{\boldmath ${\cal C}$}_x\!\cdot\!
\mbox{\boldmath ${\cal P}$}-P_0^x\right)
\nonumber\\
&\times&
\,\,\delta\!\left(
\mbox{\boldmath ${\cal C}$}_y\!\cdot\!
\mbox{\boldmath ${\cal P}$}-P_0^y\right)
\,\,\delta\!\left(
\mbox{\boldmath ${\cal C}$}_z\!\cdot\!
\mbox{\boldmath ${\cal P}$}-P_0^z\right)
\nonumber\\
&\times& \prod_i^M(2m_i)^{D/2}.
\label{4b}
\end{eqnarray}

Now we consider the following change of variables

\begin{equation}
\mbox{\boldmath ${\cal P}$}'=\mbox{\boldmath ${\cal P}$}-
\left(
\frac{\mbox{\boldmath ${\cal C}$}_x}
{|\mbox{\boldmath ${\cal C}$}_x|^2}
 P_0^x
+
\frac{\mbox{\boldmath ${\cal C}$}_y}
{|\mbox{\boldmath ${\cal C}$}_y|^2} P_0^y
+
\frac{\mbox{\boldmath ${\cal C}$}_z}
{|\mbox{\boldmath ${\cal C}$}_z|^2} P_0^z
\right),
\label{tra}
\end{equation}
which is simply a translation and has unit Jacobian. Simple algebra
leads to

\begin{eqnarray}
\Phi({\bf P}_0,E_0,M)&=& \prod_i^M(2m_i)^{D/2}
\int d^{DM}\mbox{\boldmath ${\cal P}$}'
\,\,\delta\!\left(
{\mbox{\boldmath ${\cal P}$'}^2}
-U_0\right)
\nonumber\\
&\times&
\delta\!\left(\mbox{\boldmath ${\cal
C}$}_x\!\cdot\!\mbox{\boldmath ${\cal P}$}'\right)
\,\,\delta\!\left(\mbox{\boldmath ${\cal
C}$}_y\!\cdot\!\mbox{\boldmath ${\cal P}$}'\right)
\,\,\delta\!\left(\mbox{\boldmath ${\cal
C}$}_z\!\cdot\!\mbox{\boldmath ${\cal P}$}'\right),
\label{4c}
\end{eqnarray}
where we have introduced the total internal energy

\begin{equation}
U_0=\left(E_0-\frac{{\bf P}_0^2}{2{\cal M}_0}\right),
\label{U}
\end{equation}
where ${\cal M}_0=\sum_im_i$ is the total mass.

We now consider a  second change of variables $\mbox{\boldmath  ${\cal
P}$}''={\bf  \Lambda}\!\cdot\!\mbox{\boldmath  ${\cal P}$}'$ through a
rotation ${\bf \Lambda}$ such that

\begin{eqnarray}
{\bf \Lambda}\mbox{\boldmath ${\cal C}$}_x 
&=&(1,\ldots,0,0,\ldots,0,0,\ldots,0)
\nonumber\\
{\bf \Lambda}\mbox{\boldmath ${\cal C}$}_y 
&=&(0,\ldots,0,1,\ldots,0,0,\ldots,0)
\nonumber\\
{\bf \Lambda}\mbox{\boldmath ${\cal C}$}_z 
&=&(0,\ldots,0,0,\ldots,0,1,\ldots,0).
\label{defis2}
\end{eqnarray}
It is always possible to find a  matrix ${\bf \Lambda}$ that satisfies
Eqns. (\ref{defis2}).   For example, consider  a block diagonal matrix
made of  three identical blocks of  size $M\times M$. Then assume that
each block is the same  orthogonal matrix which transforms the  vector
$((2m_1)^{1/2},(2m_2)^{1/2},\ldots,(2m_M)^{1/2})$     into    $(2{\cal
M}_0)^{1/2}(1,0,\ldots,0)$.   After the   rotation  (which   has  unit
Jacobian  and leaves the modulus of  a vector  invariant) the integral
(\ref{4c}) becomes

\begin{eqnarray}
\Phi&=&
\prod_i^M(2m_i)^{D/2}
\int d^{DM}\mbox{\boldmath ${\cal P}$}''
\delta\!\left(\mbox{\boldmath ${\cal P}$}''^2-U_0\right)
\nonumber\\
&\times&
\,\,\delta(p_1''^x)
\,\,\delta(p_1''^y)
\,\,\delta(p_1''^z)
\nonumber\\
&=&\prod_i^M(2m_i)^{D/2}
\int d^{D(M-1)}\mbox{\boldmath ${\cal P}$}''
\delta\!\left(\mbox{\boldmath ${\cal P}$}''^2-U_0\right).
\label{10}
\end{eqnarray}

We compute now the  integral  over the  sphere  in Eqn. (\ref{10})  by
using that the integral of an arbitrary function $F({\bf x}) = f(|{\bf
x}|) $ that depends on ${\bf x}$ only through  its modulus $|{\bf x}|$
can be computed by changing to polar coordinates
\begin{equation}
\int F({\bf x})d^M{\bf x}=\omega_M\int_0^{\infty} f(r) r^{M-1}dr.
\label{0}
\end{equation}
The numerical factor $\omega_M$, which comes from the integration of
the angles, can be computed by considering the special case when
$f(r)$ is a Gaussian. The result is
\begin{equation}
\omega_M = 2 \frac{\pi^{M/2}}{\Gamma(M/2)}
\label{1b}
\end{equation}
By using Eqn. (\ref{0}), Eqn. (\ref{10}) becomes

\begin{eqnarray}
\Phi({\bf P}_0,E_0,M)
&=& \prod_i^M(2m_i)^{D/2}\omega_{D(M-1)}
\nonumber\\
&\times&\int dp \,\,p^{D(M-1)-1}
\,\,\delta\!\left(p^2-U\right).
\label{10b}
\end{eqnarray}
We need now the property

\begin{equation}
\delta(f(x))=\sum_i\frac{\delta(x-x_i)}{|f'(x_i)|}
\end{equation}
where $x_i$ are the zeros of $f(x_i)=0$. For the case of
Eqn. (\ref{10b}) we have $f(x)=x^2-U$, $x_i =\pm (U)^{1/2}$ and
$f'(x)=2x$. Therefore,

\begin{equation}
\Phi({\bf P}_0,E_0,M) = \frac{1}{2}\omega_{D(M-1)}
U_0^{\frac{D(M-1)-2}{2}}\prod_i^M(2m_i)^{D/2}.
\label{11}
\end{equation}

\section{Appendix: van der Waals and hard core models}
\label{vdwhc}

The particular functional  form of  $s^{\rm hc}(\epsilon,\rho)$ cannot
be computed from the  microscopic analysis  presented in the  previous
appendix. The approach taken in  this  paper is that this  fundamental
thermodynamic equation (in the sense of Callen \cite{Callen}) is known
either  from  empirical sources or  by   suitable  modelling. In  this
appendix we summarize the results for the  fundamental equation of van
der Waals model and for its corresponding hard core model.

The  van  der  Waals  model  can   be  defined through   the following
fundamental equation,   which  relates the  entropy  density  with the
internal energy density $\epsilon$ and number density $n$,
\begin{equation}
s^{\rm vdW}(\epsilon,n) = \frac{D+2}{2}k_B n -
k_B n \ln\left(\frac{\left[\Lambda^{\rm vdW}(\epsilon,n)\right]^D n}
{1-n b}\right).
\end{equation}
The number  density is $n=\rho/m_0$ where  $\rho$ is the  mass density
and $m_0$ is the mass of a molecule. The thermal wavelength is defined
by
\begin{equation}
\Lambda^{\rm vdW}(\epsilon,n) = 
\frac{h}{(2\pi m_0k_B{\cal T}^{\rm vdW}(\epsilon,n))^{1/2}},
\end{equation}
and we have introduced the function
\begin{equation}
{\cal T}^{\rm vdW}(\epsilon,n)=\frac{2}{D}\frac{\epsilon+an^2}{k_Bn}.
\end{equation}

The two equations of  state  are the  derivatives of the  entropy with
respect to each variable
\begin{eqnarray}
\frac{\partial s^{\rm vdW}}{\partial \epsilon}
&=&\frac{1}{T^{\rm vdW}(\epsilon,n)},
\nonumber\\
\frac{\partial s^{\rm vdW}}{\partial n}
&=&-\frac{\mu^{\rm vdW}(\epsilon,n)}{T^{\rm vdW}(\epsilon,n)},
\end{eqnarray}
where $T^{\rm vdW}$ is the temperature and $\mu^{\rm vdW}$ is 
the chemical potential. Straightforward calculations lead to
\begin{eqnarray}
T^{\rm vdW}(\epsilon,n) &=&{\cal T}^{\rm vdW}(\epsilon,n),
\nonumber\\
\mu^{\rm vdW}(\epsilon,n) &=&
k_BT^{\rm vdW} \left(
\ln\left(\frac{n[\Lambda^{\rm vdW}]^D}{1-nb}\right)+\frac{nb}{1-nb}\right)
-2an.
\end{eqnarray}
The third equation  of state can be obtained  from  the Euler equation
$P= Ts -\epsilon +\mu n$, which leads to

\begin{equation}
P^{\rm vdW}(\epsilon,n) = n\frac{k_BT^{\rm vdW}(\epsilon,n)}{1-nb}-an^2.
\end{equation}

The hard core model that corresponds to the van der Waals model can be
defined as the model that results from taking the attractive parameter
$a=0$ in the van  der Waals model.  This model  is essentially the gas
ideal model but with excluded volume effects.  In this way, one obtains

\begin{equation}
s^{\rm hc}(\epsilon,n) = \frac{D+2}{2}k_B n -
k_B n \ln\left(\frac{\Lambda^{{\rm hc} D}(\epsilon,n) n}{1-n b}\right),
\end{equation}
where now
\begin{eqnarray}
\Lambda^{\rm hc}(\epsilon,n) &= &
\frac{h}{(2\pi m_0k_B{\cal T}^{\rm hc}(\epsilon,n))^{1/2}},
\nonumber\\
{\cal T}^{\rm hc}(\epsilon,n)&=&\frac{2}{D}\frac{\epsilon}{k_Bn}.
\end{eqnarray}
The equations of  state  are now
\begin{eqnarray}
T^{\rm hc}(\epsilon,n) &=&{\cal T}^{\rm hc}(\epsilon,n),
\nonumber\\
\mu^{\rm hc}(\epsilon,n) &=&
k_BT^{\rm vdW} \left(
\ln\left(\frac{n[\Lambda^{\rm hc}]^D}{1-nb}\right)+\frac{nb}{1-nb}\right),
\nonumber\\
P^{\rm hc}(\epsilon,n) &=&n \frac{k_BT^{\rm hc}(\epsilon,n)}{1-nb}.
\end{eqnarray}

Note that the following functional relations hold
\begin{eqnarray}
T^{\rm hc}(u+an^2,n)&=&T^{\rm vdW}(u,n),
\nonumber\\
P^{\rm hc}(u+an^2,n)-an^2&=&P^{\rm vdW}(u,n),
\nonumber\\
\mu^{\rm hc}(u+an^2,n)-2an&=&\mu^{\rm vdW}(u,n).
\label{conect}
\end{eqnarray}

%\emulticol
\end{document}